\documentclass[acmsmall, nonacm]{acmart}

\AtBeginDocument{%
  \providecommand\BibTeX{{%
    \normalfont B\kern-0.5em{\scshape i\kern-0.25em b}\kern-0.8em\TeX}}}

\setcopyright{acmcopyright}
\copyrightyear{2021}
\acmYear{2021}
\acmDOI{10.1145/1122445.1122456}

\acmBooktitle{HATRA '21: Human Aspects of Types and Reasoning Assistants, October 19, 2021, Chicago, IL}

\usepackage{url}
\usepackage{breakurl}
\usepackage{hyperref}

\usepackage{listings}
\usepackage{tabularx}
\usepackage{xcolor}
\usepackage{booktabs}

\usepackage{listings, xcolor}

\definecolor{verylightgray}{rgb}{.97,.97,.97}

\lstdefinelanguage{Solidity}{
	keywords=[1]{anonymous, assembly, assert, balance, break, call, callcode, case, catch, class, constant, continue, constructor, contract, debugger, default, delegatecall, delete, do, else, emit, event, experimental, export, external, false, finally, for, function, gas, if, implements, import, in, indexed, instanceof, interface, internal, is, length, library, log0, log1, log2, log3, log4, memory, modifier, new, payable, pragma, private, protected, public, pure, push, require, return, returns, revert, selfdestruct, send, solidity, storage, struct, suicide, super, switch, then, this, throw, transfer, true, try, typeof, using, value, view, while, with, addmod, ecrecover, keccak256, mulmod, ripemd160, sha256, sha3}, 
	keywordstyle=[1]\color{blue}\bfseries,
	keywords=[2]{address, bool, byte, bytes, bytes1, bytes2, bytes3, bytes4, bytes5, bytes6, bytes7, bytes8, bytes9, bytes10, bytes11, bytes12, bytes13, bytes14, bytes15, bytes16, bytes17, bytes18, bytes19, bytes20, bytes21, bytes22, bytes23, bytes24, bytes25, bytes26, bytes27, bytes28, bytes29, bytes30, bytes31, bytes32, enum, int, int8, int16, int24, int32, int40, int48, int56, int64, int72, int80, int88, int96, int104, int112, int120, int128, int136, int144, int152, int160, int168, int176, int184, int192, int200, int208, int216, int224, int232, int240, int248, int256, mapping, string, uint, uint8, uint16, uint24, uint32, uint40, uint48, uint56, uint64, uint72, uint80, uint88, uint96, uint104, uint112, uint120, uint128, uint136, uint144, uint152, uint160, uint168, uint176, uint184, uint192, uint200, uint208, uint216, uint224, uint232, uint240, uint248, uint256, var, void, ether, finney, szabo, wei, days, hours, minutes, seconds, weeks, years},	
	keywordstyle=[2]\color{teal}\bfseries,
	keywords=[3]{block, blockhash, coinbase, difficulty, gaslimit, number, timestamp, msg, data, gas, sender, sig, value, now, tx, gasprice, origin},	
	keywordstyle=[3]\color{violet}\bfseries,
	identifierstyle=\color{black},
	sensitive=false,
	comment=[l]{//},
	morecomment=[s]{/*}{*/},
	commentstyle=\color{gray}\ttfamily,
	stringstyle=\color{red}\ttfamily,
	morestring=[b]',
	morestring=[b]"
}

\begin{document}

\title{An Empirical Study of Protocols in Smart Contracts}

\author{Timothy Mou}
\email{tmou1@swarthmore.edu}
\affiliation{%
  \institution{Swarthmore College}
  \city{Swarthmore}
  \state{Pennsylvania}
  \country{USA}
}

\author{Michael Coblenz}
\email{mcoblenz@umd.edu}
\affiliation{%
  \institution{University of Maryland}
  \city{College Park}
  \state{Maryland}
  \country{USA}
}

\author{Jonathan Aldrich}
\email{jonathan.aldrich@cs.cmu.edu}
\affiliation{%
  \institution{Carnegie Mellon University}
  \city{Pittsburgh}
  \state{Pennsylvania}
  \country{USA}
}

\newcommand{\mjc}[1]{{\color{red}#1}}
\newcommand{\jea}[1]{{\color{red}#1}}

\begin{abstract}
Smart contracts are programs that are executed on a blockhain.
They have been used for applications in voting, decentralized finance, and supply chain management.
However, vulnerabilities in smart contracts have been abused by hackers, leading to financial losses.
Understanding state machine protocols in smart contracts has been identified as important to catching common bugs, improving documentation, and optimizing smart contracts.
We analyze Solidity smart contracts deployed on the Ethereum blockchain and study the prevalence of protocols and protocol-based bugs, as well as opportunities for gas optimizations.
\end{abstract}



\maketitle

\section{Introduction}
Smart contracts are general-purpose programs that are run on a blockchain and maintain state stored on a blockchain’s ledger. 
They have been used in a variety of applications, including voting, decentralized finance, and supply chain management \cite{ibm_2019, 10.1145/3173574.3174032}. 
However, vulnerabilities in smart contracts can be exploited and have serious consequences, such as the well-known DAO hack in which over \$50 million worth of cryptocurrency was stolen \cite{popper}.
Smart contracts on Ethereum cannot be modified once they are deployed, making bugs difficult to fix retroactively.
In addition, on Ethereum, the most common platform for smart contracts, users must pay \textit{gas} in order to run transactions, which is used to reward miners and prevent non-terminating transactions. High gas prices can make interacting with smart contracts prohibitively expensive, limiting their usefulness \cite{rozen_2021}.

Smart contracts have been observed to often follow \textit{protocols} that can be represented by state machines \cite{ethereum_foundation_2021}.
Protocols specify that a contract's functions must be called in a particular order.
A contract may have multiple abstract states that it transitions between, where each abstract state determines which functions can be called in that particular state.
For instance, consider a wallet contract whose public functions are shown in Figure \ref{fig:lockEtherPay}.
Its purpose is to transfer its token balance to a beneficiary after a fixed amount of time.
The wallet begins in an initial state, and to initiate the transfer, the owner of the wallet must call \texttt{lock()} to lock the wallet for a set period of time.
Only after \texttt{lock()} has been called and the time has elapsed can this state can \texttt{release()} be called to transfer the wallet's balance to the beneficiary.

\begin{figure}
\begin{center}
\begin{lstlisting}[language=Solidity,frame=single,basicstyle=\ttfamily\scriptsize]
function lock() public onlyOwner returns (bool) {
  require(!isLocked);
  require(tokenBalance() > 0);
  start_time = now;
  end_time = start_time.add(fifty_two_weeks);
  isLocked = true;
}

function lockOver() constant public returns (bool) {
  uint256 current_time = now;
  return current_time > end_time;
}

function release() onlyOwner public {
  require(isLocked);
  require(!isReleased);
  require(lockOver());
  uint256 token_amount = tokenBalance();
  token_reward.transfer(beneficiary, token_amount);
  emit TokenReleased(beneficiary, token_amount);
  isReleased = true;
}
\end{lstlisting}
\caption{The main functions in a wallet contract. The \texttt{onlyOwner} modifier on the \texttt{lock} and \texttt{release} functions allows only the owner of the wallet contract to call them.}
\label{fig:lockEtherPay}
\end{center}
\end{figure}

The concept of typestate \cite{typestate} allows one to specify the set of valid operations which can be called on a contract, which may change as the contract transitions to different states.
New programming languages for smart contracts, including Obsidian \cite{10.1145/3417516} and Flint \cite{10.1145/3191697.3213790}, have proposed that typestate can be used to express protocols more explicitly and statically check adherence to protocols.
One benefit of such systems is that they allow developers to specify functions and fields that are only in scope in certain states.
Failure to change or reset fields between state transitions can lead to bugs that are difficult to track down.
In addition, accessing storage on the blockchain is costly, and using typestate may allow contracts to lower their gas usage by clearing fields or avoiding unnecessary writes to storage.

We conducted an empirical study of Solidity smart contracts to demonstrate that protocols are commonly used in smart contracts and that typestate can help contracts avoid bugs and use less gas.
We collected a sample of 100 smart contracts and determined that 37 of them defined protocols.
We wrote a static analysis tool, StateOptimizationDetector, to find opportunities to lower the cost of transactions by taking advantage of Ethereum's gas refund policy.
We ran StateOptimizationDetector on a sample of 1,000 contracts and found that 80 (0.8\%) of them could use this optimization, with a false positive rate of 28.6\%.

We investigated the prevalence of one access control bug, which allows previous owners of a contract to unexpectedly reclaim ownership, to show that the incorrect implementation of protocols can create vulnerabilities.
We wrote a static analysis tool, AccessLockDetector, to identify contracts which could have bugs in the implementation of their access control mechanisms.
We ran AccessLockDetector on a sample of 4,500 contracts and identified 36 (0.8\%) of them via manual examination as vulnerable to this exploit, with a false positive rate of 34.5\%.
We observed a high level of similarity among the identified vulnerable contracts, suggesting that many of them originated from a common source. 
We hypothesize that language features that allow protocols to be expressed more directly can help contract writers avoid these bugs or catch them before deployment.

\section{Methodology}

To inform our understanding of how smart contracts define and use protocols, we first tried to identify contracts with protocols by looking for patterns in contracts' source code.
We built a detector using the Slither static analysis tool \cite{Feist_2019}, named ProtocolDetector, that identifies contracts with protocols by searching for reverting execution paths in functions that resulted from reading the contract's fields. 
The motivation behind this is that when a function in a protocol-defining contract is called, the contract will often first check that it is in an appropriate abstract state, which is typically represented by one or more fields.
Reverting paths that result from reading fields often indicate that the protocol has been violated. 
Similar patterns have been used to search for evidence of object protocols in Java \cite{10.1007/978-3-642-22655-7_2}.
For example, in Figure \ref{fig:lockEtherPay}, the function \texttt{release} will revert if the values of the fields \texttt{isLocked} and \texttt{isReleased} are not \texttt{true} and \texttt{false}, respectively.
This indicates that these fields represent the contract's abstract state, and that this contract defines a protocol.

We incorporated several heuristics specific to Solidity to help the detector find common patterns indicative of protocols. For example, Solidity allows the definition of \textit{modifiers}, which can be attached to function declarations to insert code that is run before or after the main body of the function.
When checking a function for state patterns, the detector will also check the body of any of its modifiers, since they are commonly used to ensure that the contract is in a valid state before executing the function.

\subsection{Finding protocol-based optimizations}
We then identified a method to lower the gas cost of some transactions using Ethereum's refund mechanism.
When a storage value is reset (set to zero from non-zero), a refund of (up to) 15,000 gas is given back to the sender of the transaction \cite{wood2014ethereum}.
Contracts with certain types of protocols may be able to use the refund to lower the cost of some of their transactions.
When a contract transitions to a new state, some of its fields may no longer be used for the remaining lifetime of the contract. 
These fields could therefore be reset during this state transition to take advantage of the refund, lowering the overall gas cost. 
This optimization encourages the clearing of unused fields, reducing the amount of unneeded state that is kept on the Ethereum blockchain \cite{moosavi2021lissy}.
See Figure \ref{fig:Deactivation} for an example case of this optimization.
Here, the function \texttt{addToX} is ``deactivated'' by calling \texttt{stop()}.
Since the field \texttt{amountToAdd} is only used in the function \texttt{addToX}, after \texttt{stop()} is called, \texttt{amountToAdd} can safely be reset without affecting any observable behavior of the contract.
It should be noted that a recent proposal to Ethereum that is included in the London hard fork, EIP-3529 \cite{eip_3529}, reduces the amount of gas that is refunded by clearing storage, making this optimization no longer viable after the proposal is implemented.
However, this optimization is still applicable to earlier versions of Ethereum.

\begin{figure}
\begin{lstlisting}[language=Solidity,frame=single,basicstyle=\ttfamily\scriptsize]
contract Deactivation {
    bool public stopped;
    uint public x;
    uint private amountToAdd;

    function stop() public {
        require(!stopped);
        stopped = true;
        amountToAdd = 0; //This field can safely be reset here.
    }

    function addToX() public {
        require(!stopped);
        amountToAdd++;
        x += amountToAdd;
    }
}
\end{lstlisting}
\caption{An example of a protocol-based optimization.}
\label{fig:Deactivation}
\end{figure}

To investigate how useful this optimization was, we identified two categories of protocols (described in Section \ref{OptimizationResults}) that indicated that this optimization was applicable. 
Then we built a detector, named StateOptimizationDetector, to find contracts with these protocols.
This detector infers the contract's transition graph between different abstract states by identifying the set of states a contract can be in before and directly after calling a function. 
The detector uses the state transition graph to identify fields that will never be used after a state transition, meaning that it will never be read from or written to in any function that can be called in the current state or any possible future state. 

We manually examined each candidate contract reported by the StateOptimizationDetector.
During this process, we checked that each candidate contract had one of the two protocol patterns, and that each identified field could safely be reset without affecting the contract's expected behavior.
One issue we ran into was that some fields are declared public, meaning that a public getter function for the field is automatically generated and added to the contract's interface.
Thus, resetting this field could affect the contract's clients that access this getter function after the state transition, even if it does not affect the behavior of the contract itself.
However, the fact that this field is no longer used may also indicate that this field is conceptually no longer in scope during this state.
Whether to count such a field depends on the specifics of the contract, and when manually examining the results, we only counted the fields that we deemed safe to reset.
For example, in some contracts, we noted that fields would not be used within the contract, but were simply used to maintain some status that would be useful to other contracts. 
An example is shown in Figure \ref{fig:MintableToken}.
Such contracts were discarded during the manual review process.

\begin{figure}
\begin{lstlisting}[language=Solidity,frame=single,basicstyle=\ttfamily\scriptsize]
contract MintableToken is StandardToken, Ownable {
  uint public totalSupply;
  bool public mintingFinished = false;

  modifier canMint() {
    require(!mintingFinished);
    _;
  }

  function mint(address _to, uint256 _amount) onlyOwner canMint returns (bool) {
    totalSupply = totalSupply.add(_amount);
    balances[_to] = balances[_to].add(_amount);
    return true;
  }

  function finishMinting() onlyOwner returns (bool) {
    mintingFinished = true;
    return true;
  }
}
\end{lstlisting}
\caption{A token contract with a field \texttt{totalSupply} that maintains the total number of tokens. After \texttt{finishMinting()} is called, the value of \texttt{totalSupply} can no longer change, but it should not be reset, since the total number of tokens should still be available even after no more tokens can be minted.}
\label{fig:MintableToken}
\end{figure}

\subsection{Finding bugs in access control contracts}

\begin{figure}
\begin{lstlisting}[language=Solidity,frame=single,basicstyle=\ttfamily\scriptsize]
contract Ownable is Context {
    address private _owner;
    address private _previousOwner;
    uint256 private _lockTime;

    ...

    /**
    * @dev Throws if called by any account other than the owner.
    */
    modifier onlyOwner() {
        require(_owner == _msgSender(), "Ownable: caller is not the owner");
        _;
    }

    /**
    * @dev Leaves the contract without owner. It will not be possible to call
    * `onlyOwner` functions anymore. Can only be called by the current owner.
    *
    * NOTE: Renouncing ownership will leave the contract without an owner,
    * thereby removing any functionality that is only available to the owner.
    */
    function renounceOwnership() public virtual onlyOwner {
        emit OwnershipTransferred(_owner, address(0));
        _owner = address(0);
    }

    /**
    * @dev Transfers ownership of the contract to a new account (`newOwner`).
    * Can only be called by the current owner.
    */
    function transferOwnership(address newOwner) public virtual onlyOwner {
        require(newOwner != address(0), "Ownable: new owner is the zero address");
        emit OwnershipTransferred(_owner, newOwner);
        _owner = newOwner;
    }

    function geUnlockTime() public view returns (uint256) {
        return _lockTime;
    }

    //Locks the contract for owner for the amount of time provided
    function lock(uint256 time) public virtual onlyOwner {
        _previousOwner = _owner;
        _owner = address(0);
        _lockTime = now + time;
        emit OwnershipTransferred(_owner, address(0));
    }
    
    //Unlocks the contract for owner when _lockTime is exceeds
    function unlock() public virtual {
        require(_previousOwner == msg.sender, "You don't have permission to unlock");
        require(now > _lockTime , "Contract is locked until 7 days");
        emit OwnershipTransferred(_owner, _previousOwner);
        _owner = _previousOwner;
    }
}  
\end{lstlisting}
\caption{A Ownable contract with a temporary lock mechanism (lines 42-56).
Note that the \texttt{unlock} function allows ownership to be reclaimed after it has been renounced or transferred to a new owner.}
\label{fig:Ownable}
\end{figure}

Access control mechanisms are used in smart contracts to restrict access to certain administrative functions.
An example of an access control contract is the \texttt{Ownable} contract shown in Fig. \ref{fig:Ownable}, which defines an \texttt{onlyOwner} modifier that can be attached to functions, which allows only the owner of a contract to call them.
Contracts can inherit from the \texttt{Ownable} contract and then use the \texttt{onlyOwner} modifier to restrict access to certain functions.

Access control contracts also commonly have the ability to transfer ownership to another address and renounce ownership (preventing anyone from calling \texttt{onlyOwner} functions).
Some access control contracts also have a temporary lock feature, which allows the owner to temporarily relinquish ownership, preventing themselves and anyone else from calling \texttt{onlyOwner} functions for a set period of time.
During this time, users of a contract can be assured that \texttt{onlyOwner} functions cannot be called.
A contract with both a temporary lock and the ability to renounce ownership can be modeled with a protocol with three abstract states, Locked, Unlocked, and Renounced. 

However, we observed that many implementations of this temporary lock feature violate this protocol and create a vulnerability in the expected behavior of renouncing and transferring ownership.
As shown in Table \ref{tab:RenounceBugSteps}, after the function \texttt{unlock} is called once, it can continue to be called even after the owner has renounced or transferred ownership, since the value of \texttt{\_previousOwner} is never reset.
This interaction may affect users of a contract who make decisions based on the assumption that \texttt{onlyOwner} functions cannot be called after a contract is renounced.
For instance, a malicious owner of a token contract could renounce ownership to convince prosepective buyers that the price of the token is fixed, then use this exploit to reclaim ownership and raise the price of the token.

\begin{table}[]
    \centering
    \caption{Steps to exploit a vulnerability in the \texttt{Ownable} contract in Fig. \ref{fig:Ownable}, where the owner of a contract can take back ownership after they have already renounced ownership.}
    \begin{tabularx}{\linewidth}{X c c c}\toprule
         &  \texttt{\_owner} & \texttt{\_previousOwner} & Contract state \\ \midrule
         Contract is initialized by an address \textit{addr}. \texttt{\_owner} is set to \textit{addr}. & \textit{addr} & 0x0 & Unlocked \\ 
         Owner calls \texttt{lock(0)}. & 0x0 & \textit{addr} & Locked \\
         Owner calls \texttt{unlock()}. & \textit{addr} & \textit{addr} & Unlocked \\
         Owner calls \texttt{renounceOwnership()}. & 0x0 & \textit{addr} & Renounced \\
         Owner calls \texttt{unlock()}. & \textit{addr} & \textit{addr} & Unlocked \\ \bottomrule
    \end{tabularx}
    
    \label{tab:RenounceBugSteps}
\end{table}

To investigate how commonly this bug occurs, we built a detector, AccessLockDetector, which is designed to identify access control contracts which define temporary lock mechanisms.
To find access control contracts, it looks for contracts that have an ``owner'' address field, which represents the only address that can call restricted functions, and define an \texttt{onlyOwner} modifier, which modifies functions so that they can only be called by the contract's owner. 
Then, it looks for a ``locking'' function in which the owner address is first assigned to another field and then reset. This pattern indicates that owner's access has been disabled, and that the owner address has been saved to another field so access can be re-enabled later. 
We observed that contracts that implemented this temporary lock mechanism frequently contain a bug that breaks the expected behavior of renouncing and transferring ownership.

\section{Results}

\subsection{Prevalence of protocol-defining contracts}

We ran the ProtocolDetector on a random sample of 100 contracts from the SmartBugs Wild dataset \cite{SmartBugs}, a collection of over 47,398 unique Solidity contracts.
ProtocolDetector identified 71 protocols in 37 contracts.
We manually reviewed each reported protocol and categorized them based on each protocol's intended use and method of transitioning between states.
We determined that 2 (2.8\%) of the protocols found were false positives. However, in both cases, the contract defined other protocols, so in total, 37 of the contracts defined protocols.
The categorization of protocols, which is based of a categorization of Java protocols in \cite{10.1007/978-3-642-22655-7_2}, is shown in Table \ref{tab:results1}.

\begin{table}
  \caption{Manual categorization of 69 protocols identified with the ProtocolDetector from a random sample of 100 contracts.}
  \label{tab:results1}
  \begin{tabularx}{\linewidth}{llX}
    \toprule
    Category & Count & Description\\
    \midrule
    Toggle              & 21 & The owner of a contract can enable/disable functionality at will. \\
    Activation          & 12 & Some functions become eventually enabled. \\
    Redundant Operation & 12 & Prevents a function from being called twice. \\
    Deactivation        & 10 & Some functions become eventually disabled. \\
    Threshold           & 6  & A function is enabled once a counter reaches a certain value. \\
    Interval            & 3  & A function is enabled during a specified interval of time. \\
    Lifecycle           & 3  & The contract transitions through various states modeling the lifetime of the contract. \\
    Other               & 2  &  \\
  \bottomrule
\end{tabularx}
\end{table}

\label{OptimizationResults}
\subsection{Protocol-based optimizations}

From the categories of protocols that we had observed, two of them indicated contracts which could possibly be optimized by resetting fields.
\textit{Deactivation} protocols have two states that are represented by a boolean field and have a set of functions that are initially enabled. Once the boolean value changes, the functions are permanently disabled.
\textit{Lifecycle} protocols typically have a set of states through which the contract monotonically transitions through, eventually ending in some terminal state. 
Contracts which have either Deactivation or Lifecycle protocols permanently lose functionality as the contract transitions, which may lead to fields which are no longer used and can be safely reset.

We ran the StateOptimizationDetector on a random sample of 1,000 contracts from the SmartBugs Wild dataset.
The StateOptimizationDetector reported 112 of these as candidates for optimization, 80 of which we determined were true positives.

\begin{table}
  \caption{Results of running the StateOptimizationDetector on a random sample of 1,000 contracts.}
  \label{tab:results2}
  \begin{tabular}{lcc}
    \toprule
    Category & Candidates & Optimizable \\
    \midrule
    Deactivation & 96 & 73 \\
    Lifecycle    & 16 & 7  \\
  \midrule
    Total        & 112 (11.2\%) & 80 (8.0\%) \\
  \bottomrule
\end{tabular}
\end{table}

\subsection{Protocol-based access control bugs}

We ran the AccessLockDetector on a random sample of 4,500 contracts from the Smart Contract Sanctuary \cite{smart_contract_sanctuary}, a collection of over 99,582 smart contracts deployed on the main Ethereum blockchain, of which the AccessLockDetector flagged 55 of these contracts.
We used the Smart Contract Sanctuary dataset because it is updated daily, and we observed that the access control bug we were looking for only appeared in contracts that were deployed within the past year.
We chose a sample of 4,500 contracts so that all of the contracts flagged by the detector could be reviewed by hand.
We then manually examined each candidate contract, checking if it was an access control contract that defined a temporary lock as well as functions for renouncing ownership. Then we checked if its implementation allowed the \texttt{unlock} function to be called when the contract was already unlocked, creating the vulnerability where ownership could be reclaimed after renouncing ownership.
After examining the positive results, we determined that 36 of them were access control contracts that defined a temporary lock feature and had this vulnerability.

All of the 36 true positives followed the same structure as in Figure \ref{fig:Ownable} in defining the \texttt{lock} and \texttt{unlock} functions.
34 of these contracts contained a misleading error message in the \texttt{unlock} function (line 53 of Figure \ref{fig:Ownable}), and 33 of them contained the same typo in a function name (line 38 of Figure \ref{fig:Ownable}).
In addition, all of the contracts passed an incorrect argument to the \texttt{OwnershipTransferred} event in the \texttt{lock} function (line 47 of Figure \ref{fig:Ownable}; the first argument should be \texttt{\_previousOwner}, since \texttt{\_owner} has already been reset). Although events are not used by other smart contracts, this could provide incorrect information to applications outside the blockchain that are subscribed to this event.

\section{Related Work}
Previous studies have employed static analysis and symbolic execution tools to analyze bugs and security vulnerabilities in smart contracts.
We built our detectors using Slither \cite{Feist_2019}, an extensible static analysis tool which analyses Solidity source code and converts it into a convenient intermediate representation.
Oyente \cite{Oyente} is a symbolic execution tool that finds common bugs involving issues with transaction-ordering dependence and reentrancy attacks.
Smartcheck \cite{Smartcheck} runs a syntactical analysis on Solidity source code to detect a number of code issues and poor practices.

A high level of code cloning and duplication among smart contracts has been shown through various studies and smart contract datasets.
In the SmartBugs Wild dataset \cite{SmartBugs}, only 47,398 out of 972,855 (4.8\%) verified  contracts (those with source code available) were unique. 
In a dataset of 22,493 smart contracts used for \cite{kalra2018zeus}, only 1,524 (6.8\%) of contracts were unique.
A study on code cloning in smart contracts found that only 20.8\% of verified contracts were not a clone of any other verified contract \cite{CodeCloning}.

Gasper \cite{Gasper} is a tool that locates gas-costly patterns by analyzing contracts' bytecodes. It reduces the cost of transactions by eliminating dead code and rewriting expensive loop operations.
We are not aware of any tools which optimize gas usage by taking advantage of Ethereum's gas refund policy.

\section{Discussion and Future Work}
We have shown that protocols appear relatively frequently in Solidity smart contracts, with 37\% of contracts defining protocols.
For comparison, it is estimated that 7.2\% of types in Java programs define protocols \cite{10.1007/978-3-642-22655-7_2}.
Therefore, we believe the implementation of protocols is an important consideration for the design of languages for smart contracts.
Moreover, a number of categories of protocols allow for gas optimizations, where fields can safely be reset to reduce the cost of transactions.

We have demonstrated that approximately 0.8\% of deployed Ethereum contracts contain a bug in their access control mechanisms that stems from a failure to accurately model a contract's protocol.
We believe that introducing language features such as typestate could help developers avoid writing these bugs like this in the first place.
Analyzing the \texttt{Ownable} contract from Figure \ref{fig:Ownable}, we notice that the vulnerability arises from the fact that the \texttt{unlock} function can be called even when the contract is not locked because the value of \texttt{\_previousOwner} is never reset.
In a language such as Obsidian, one can describe a contract as having multiple abstract states, and functions and fields can be specified as being in scope only in certain states.
This allows the expected behavior of the \texttt{Ownable} to be expressed more easily: the \texttt{Ownable} contract has three states, Locked, Unlocked, and Renounced, and the \texttt{unlock} function is only in scope in the Locked state.
In addition, one can avoid having to define a \texttt{\_previousOwner} field altogether simply by specifying the \texttt{onlyOwner} modifier to only be in scope in the Unlocked state, preventing \texttt{onlyOwner} functions from being called if the contract is in the Locked or Renounced states.

We found that many of the vulnerable contracts flagged by the AccessLockDetector had nearly identical code in their access control functions and contained the same mistakes in function names and error messages.
This supports previous results about the high level of code cloning and duplication between smart contracts.
We hypothesize that these mistakes indicate that these contracts were copied from a single source, rather than created independently.
This suggests that the practice of copying code from other contracts may introduce vulnerabilities and propagate the spread of vulnerable contracts.

In future work, we would like to investigate other ways protocols can be used to reduce gas usage. For instance, if two fields are used in different states, it may be possible to use the same storage for both fields, eliminating the cost of reserving an extra storage spot.
We would also be interested in studying protocol-related asset bugs, where a contract's assets are lost or can no longer be accessed due to a state transition.
We would also like to reduce the false positive rates on our StateOptimizationDetector and AccessLockDetector to make them more useful to Solidity developers.
We hope this work can inform and motivative future research into the design of languages and type systems for smart contracts.

\bibliographystyle{ACM-Reference-Format}
\bibliography{bibliography}

\end{document}